\documentclass[article, preprint, sort&compress]{revtex4}

\usepackage{color}
\usepackage{caption}
\usepackage{graphicx}
\usepackage{amsmath}
\usepackage{amstext}
\usepackage{subfigure}
\usepackage{units}
\usepackage{epstopdf}

\newcommand{\ket}[1]{| #1 \rangle}

\begin{document}

\title{Photon-pair generation in photonic crystal fibre with a 1.5\,GHz modelocked VECSEL}
\author{Oliver J. Morris$^{1, \dagger}$, Robert J. A. Francis-Jones$^{2, \dagger}$, Keith G. Wilcox$^3$, Anne C. Tropper$^1$, Peter J. Mosley$^2$}
\email{p.mosley@bath.ac.uk}

\affiliation{$^1$School of Physics and Astronomy, University of Southampton, Southampton, SO17 1BJ, UK \\
$^2$Centre for Photonics and Photonic Materials, Department of Physics, University of Bath, Bath, BA2 7AY, UK \\
$^3$Division of Engineering, Physics and Mathematics, University of Dundee, Nethergate, Dundee, DD1 4HN}

\footnote{$^\dagger$These authors contributed equally.}

\begin{abstract}
Four-wave mixing (FWM) in optical fibre is a leading technique for generating high-quality photon pairs. We report the generation of photon pairs by spontaneous FWM in photonic crystal fibre pumped by a 1.5\,GHz repetition-rate vertical-external-cavity surface-emitting laser (VECSEL). The photon pairs exhibit high count rates and a coincidence-to-accidental ratio of over 80. The VECSEL's high repetition-rate, high average power, tunability, and small footprint make this an attractive source for quantum key distribution and photonic quantum-state engineering. 
\end{abstract}


\maketitle

\section{Introduction}

The drive towards photonic quantum-enhanced technologies is placing ever more stringent demands on the performance of nonclassical light sources \cite{OBrien2009Photonic-quantum-technologies, Eisaman2011Single-photon-sources-and-detectors, Pan2012Multiphoton-entanglement-and-interferometry}. However due to their size, cost, complexity, and limited generation rates it is challenging to incorporate state-of-the-art photon sources into turn-key systems. Nevertheless in recent years, huge advances have been made in producing higher-quality single photons from straightforward equipment operating at room temperature. Optical nonlinearity has been at the forefront of these efforts \cite{Soller2010Bridging-visible-and-telecom, Xiong2011Slow-light-Enhanced-Correlated, Collins2012Low-Raman-noise-correlated-photon-pair, Steinlechner2012A-high-brightness-source-of-polarization-entangled, Karpinski2012Dispersion-based-control-of-modal, Pomarico2012MHz-rate-and-efficient-synchronous, McMillan2013Two-photon-interference-between}.

By propagating a high-intensity laser pulse through a nonlinear medium, pairs of photons can be spontaneously generated, either through parametric downconversion (PDC), a three-wave mixing process that requires the presence of $\chi^\text{(2)}$ nonlinearity, or by $\chi^\text{(3)}$-mediated four-wave mixing (FWM). The highest-performance sources are usually pumped by Ti:Sapphire oscillators \cite{Mosley2008Heralded-Generation-of-Ultrafast, Ling2009Mode-expansion-and-Bragg, Eckstein2011Highly-Efficient-Single-Pass, Clark2011Intrinsically-narrowband-pair, Tanida2012Highly-indistinguishable-heralded, Jin2013Widely-tunable-single}, placing them orders of magnitude higher in both complexity and cost than the attenuated laser sources typically used in commercially-available quantum technologies. Furthermore, the repetition rate of these lasers, usually 80\,MHz or so, places a limit on the rate at which high-quality photon pairs can be delivered by a single source \cite{Christ2012Limits-on-the-deterministic-creation}. These factors present significant obstacles to implementing real-world photonic quantum-enhanced technologies.

In this paper we demonstrate photon-pair generation through four-wave mixing in a photonic crystal fibre (PCF) driven by a 1.5\,GHz modelocked tunable vertical-external-cavity surface-emitting laser (VECSEL) \cite{Morris2012A-wavelength-tunable-2-ps}. VECSELs have not previously been used for photon-pair generation but are attractive for a number of reasons \cite{Quarterman2009A-passively-mode-locked-external-cavity}. Modelocked VECSELs produce transform-limited ultrafast pulses with peak power in the correct range for photon-pair generation by FWM in fibre \cite{Wilcox20134.35-kW-peak-power}. Modelocking a VECSEL requires a compact cavity containing only three or four components providing the possibility of very small-footprint photon-pair sources that require little maintenance. The short cavity results in a high pulse repetition frequency which could allow an order of magnitude increase in the rate of pair generation relative to 80\,MHz laser systems. VECSELs are intrinsically flexible: the wavelength can be continuously tuned by an intracavity etalon \cite{Morris2012A-wavelength-tunable-2-ps}; the repetition rate can be adjusted without interrupting modelocking by translating the output coupler \cite{Wilcox2011Repetition-frequency-tunable-mode-locked-surface}; and the wavelength range and pulse duration (and hence the spectral bandwidth) can be changed by using different semiconductor substrates. Such flexibility gives great scope for optical quantum-state engineering, as discussed towards the end of this paper.

\section{Four-wave mixing in photonic crystal fibre}

Photon-pair generation by degenerate FWM involves the annihilation of two photons and the corresponding creation of a photon pair, known as the signal and idler, at equal frequency detunings above and below the pump. All that is required is a material providing $\chi^{(3)}$ nonlinearity and sufficient peak power to access the nonlinearity; hence the process is usually pumped by high-power laser pulses. While the strength of the $\chi^{(3)}$ nonlinearity is orders of magnitude smaller than that available in the equivalent $\chi^{(2)}$ pair generation process -- PDC -- the ubiquity of the $\chi^{(3)}$ nonlinearity makes photon-pair generation by spontaneous FWM possible in optical fibre, where long interaction lengths are available. This also has the advantage of generating photon pairs in a well-defined guided mode, ensuring efficient collection and straightforward interfacing with subsequent fibre components or waveguiding optical chips \cite{Sansoni2012Two-Particle-Bosonic-Fermionic-Quantum, Spring2013Boson-Sampling-on-a-Photonic, Broome2013Photonic-Boson-Sampling}.

FWM in photonic crystal fibre (PCF) is ideally placed to exploit the properties of VECSELs. Unlike many semiconductor lasers, the external cavity endows VECSELs with exceptional beam quality allowing efficient coupling into the fundamental mode of a fibre. The long interaction length and small mode area in fibre means that the peak power required for photon-pair generation is relatively low; indeed, if the peak power is too high parasitic nonlinear effects such as Raman scattering will dominate and degrade the quality of the quantum states produced. The high repetition rate of a modelocked VECSEL means that even at large average powers the pulse energy is small ($\sim$\,500\,pJ) and the peak power commensurately low, which helps to limit unwanted nonlinear effects. Finally, the dispersion of PCF can be modified during fabrication by selecting the dimensions of the photonic crystal cladding; coupled with the flexibility in bandwidth and wavelength tunability of VECSELs this gives significant control over the properties of the generated photon pairs.

When FWM is pumped by a single laser pulse (the degenerate pump case) and all fields are co-polarised and propagate in the fundamental mode of a fibre, the phasematching conditions that govern the generation of photon pairs can be written
\begin{align}
2 \omega_\text{p} - \omega_\text{s} - \omega_\text{i} = 0 \\
2 k_\text{p} - k_\text{s} - k_\text{i} - 2 \gamma P_0 = 0
\end{align}
where angular frequencies $\omega_\text{p,s,i}$ and wavenumbers $k_\text{p,s,i}$ correspond to the pump, signal, and idler respectively. $\gamma$ is the nonlinearity of the fibre and $P_0$ the peak power of the pump. The resulting quantum state is correlated in photon number but dominated by the vacuum component; it can be written:
\begin{equation}
\ket{\Psi_\text{FWM}} \propto \ket{0_\text{s}, 0_\text{i}} + \lambda \ket{1_\text{s}, 1_\text{i}} + \lambda^2 \ket{2_\text{s}, 2_\text{i}} + \cdots
\label{eq:state}
\end{equation}
where $\lambda$ is proportional to $P_0$ and the generation probability for $n$ pairs is $P(n)\propto |\lambda|^{2n}$ \cite{Chen2005Two-photon-state-generation-via-four-wave, Lin2007Photon-pair-generation-in-optical}. The $\ket{0_\text{s}, 0_\text{i}}$ term is usually eliminated by heralding (detecting one photon to announce the presence of the other) or post-selecting on a successful generation event.

In order to generate high-quality photon pairs, we wish to maximise the contribution from the $\ket{1_\text{s}, 1_\text{i}}$ component in Equation \ref{eq:state} and minimise the higher-order terms \cite{Migdall2002Tailoring-single-photon-and-multiphoton}. It is clear from Equation \ref{eq:state} that although increasing $\lambda$ by turning up the peak power $P_0$ will yield a higher pair generation rate, it will also increase the emission rate of two or more pairs simultaneously. This rapidly degrades the quality of the photon-pair source. As a result, sources are usually limited to $\lambda \ll 1$ so that the pair generation probability per pulse is below 1\%, limiting the total number of pairs that can be generated per second.

Changing the pump laser repetition rate can reconcile these conflicting demands \cite{Broome2011Reducing-multi-photon-rates}. Increasing the repetition rate by a factor $m$ reduces the peak power to $P_0/m$ and therefore $\lambda \rightarrow \lambda/m$, leading to a reduction of order $m^2$ in the relative contribution of multiple-pair emission compared to singe-pair generation. Although reducing the peak power at constant average power also reduces the rate of single pairs generated by FWM by a factor of $m$, pulsed FWM sources are anyway forced to operate far below the maximum average power that the pump laser is capable of delivering. Hence the single-pair generation rate can be restored by increasing the average pump power, and the ratio of two-pair to one-pair generation events will still be reduced by a factor of $m$. Conversely, if the pair generation probability per pulse is kept constant, increasing the repetition rate allows the pair generation rate to increase by the same factor $m$ without any impact on the ratio of two-pair to one-pair generation events.

\section{Vertical-external-cavity surface-emitting laser}

\begin{figure}[h!]
\centering
\includegraphics[width=10cm]{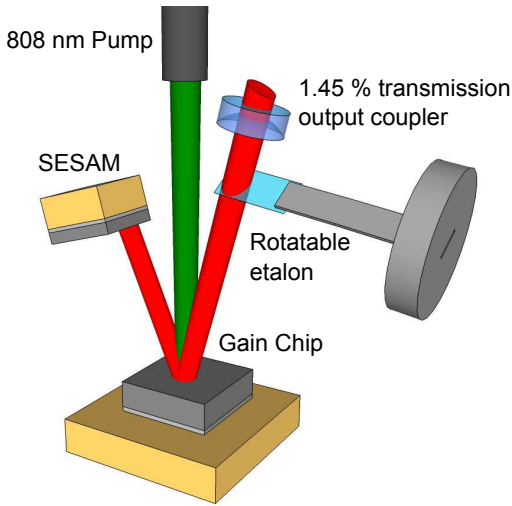}
\caption{Schematic of the laser cavity. See text for details.}
\label{fig:vecsel_cavity}
\end{figure}

The VECSEL, shown schematically in Figure \ref{fig:vecsel_cavity}, is an optically pumped, semiconductor quantum-well laser. The gain chip of the laser, fabricated by NAsP III/V GmbH (Marburg, Germany), was epitaxially grown on top of a 500\,$\mu$m thick GaAs substrate. It consisted of an InGaP cap layer, a 12$\lambda$/2 thick GaAsP active region containing 11 InGaAs quantum wells and a 22-pair AlGaAs/AlAs distributed Bragg reflector (DBR) in this order. Solid-liquid inter-diffusion bonding was  used to place a diamond heat spreader in direct thermal contact with the DBR and a chemical etch was applied to remove the substrate. The structure was then mounted diamond side down on a water-cooled copper heat-sink maintained at 18 $^\circ$C.

The VECSEL was modelocked by incorporating a Semiconductor Saturable Absorber Mirror (SESAM) into the cavity. The SESAM structure used in this work was provided by BATOP optoelectronics (SAM-1064-0.7-1ps) and had a modulation depth of 0.4\,$\%$, a non-saturable loss of 0.3\,$\%$, a saturation fluence of 130\,$\mu$J/cm$^2$ and a damage threshold of 3\,mJ/cm$^2$. The cavity was completed with a 1.45\,$\%$ transmission, 100\,mm radius of curvature (ROC) output coupler. 

Pump light is absorbed by the quantum well spacer layers creating carriers that become localised within the quantum wells. An 808\,nm fibre-coupled diode laser was used to pump the gain chip with 23\,W of light focused to a 125\,$\mu$m radius spot. In order to match the absorption of the SESAM to the emission of the gain chip, the SESAM was cooled to -10$^\circ$C using a Peltier-cooled heat-sink.

The external cavity formed by the free-standing output coupler and SESAM allows the insertion of intracavity elements. An uncoated 25\,$\mu$m-thick fused silica etalon was placed into the VECSEL cavity between the output coupler and the gain chip and held close to the Brewster angle of the s-polarised laser mode. The thickness of the etalon equates to a free spectral range (FSR) at normal incidence of 14\,nm. The incorporation of the etalon both stabilised the mode-locked operation and, by rotating the etalon about the Brewster angle, provided $\pm$4.5\,nm of wavelength selectivity around a centre wavelength of 1027\,nm. 

\begin{figure}
      \subfigure[]{
                \includegraphics[width=0.4\textwidth]{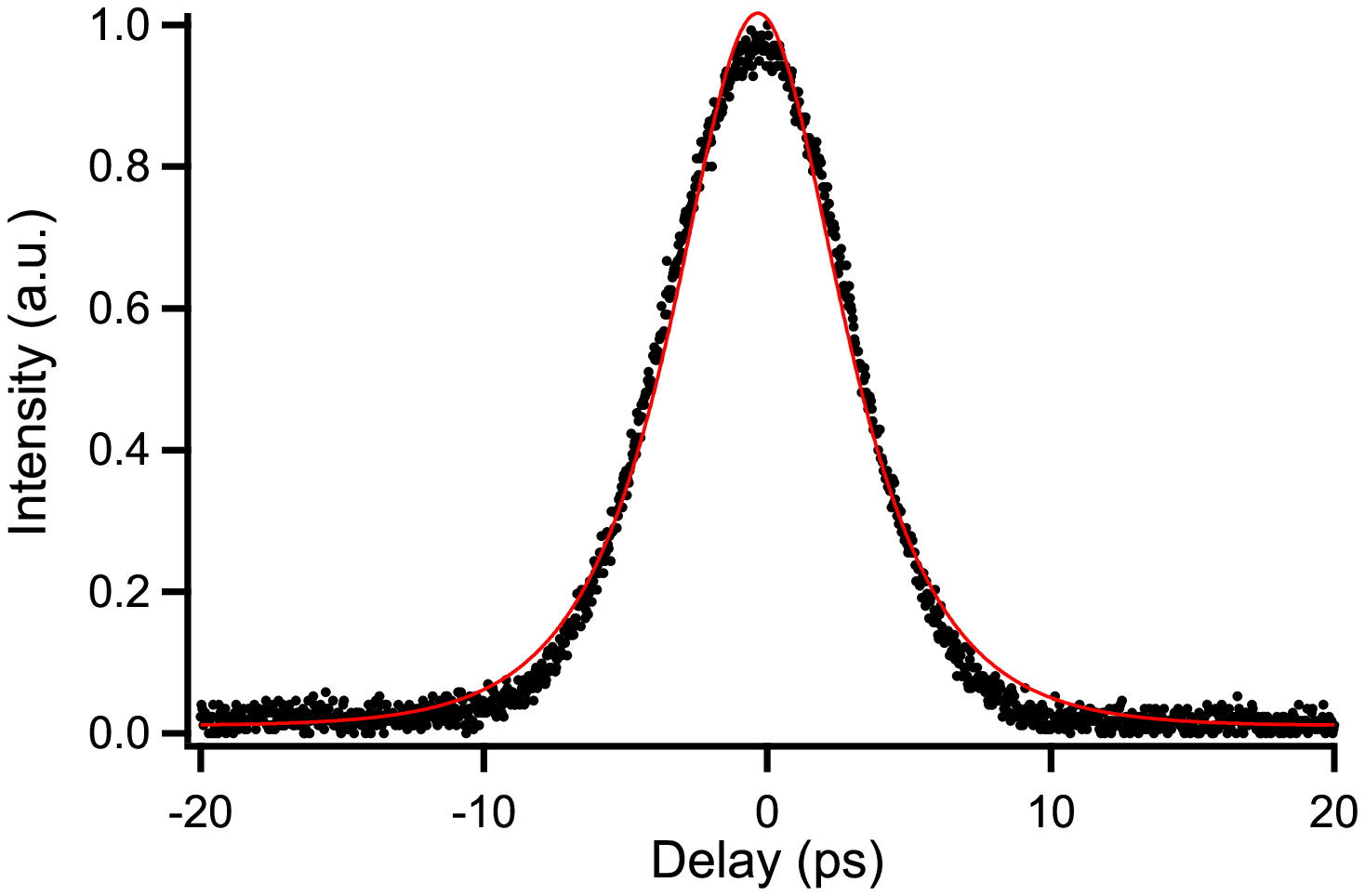}
                %\caption{Autocorrelation}
                \label{fig:autocorrelation}
       			 }
	\hspace*{\fill}
      \subfigure[]{
                \includegraphics[width=0.4\textwidth]{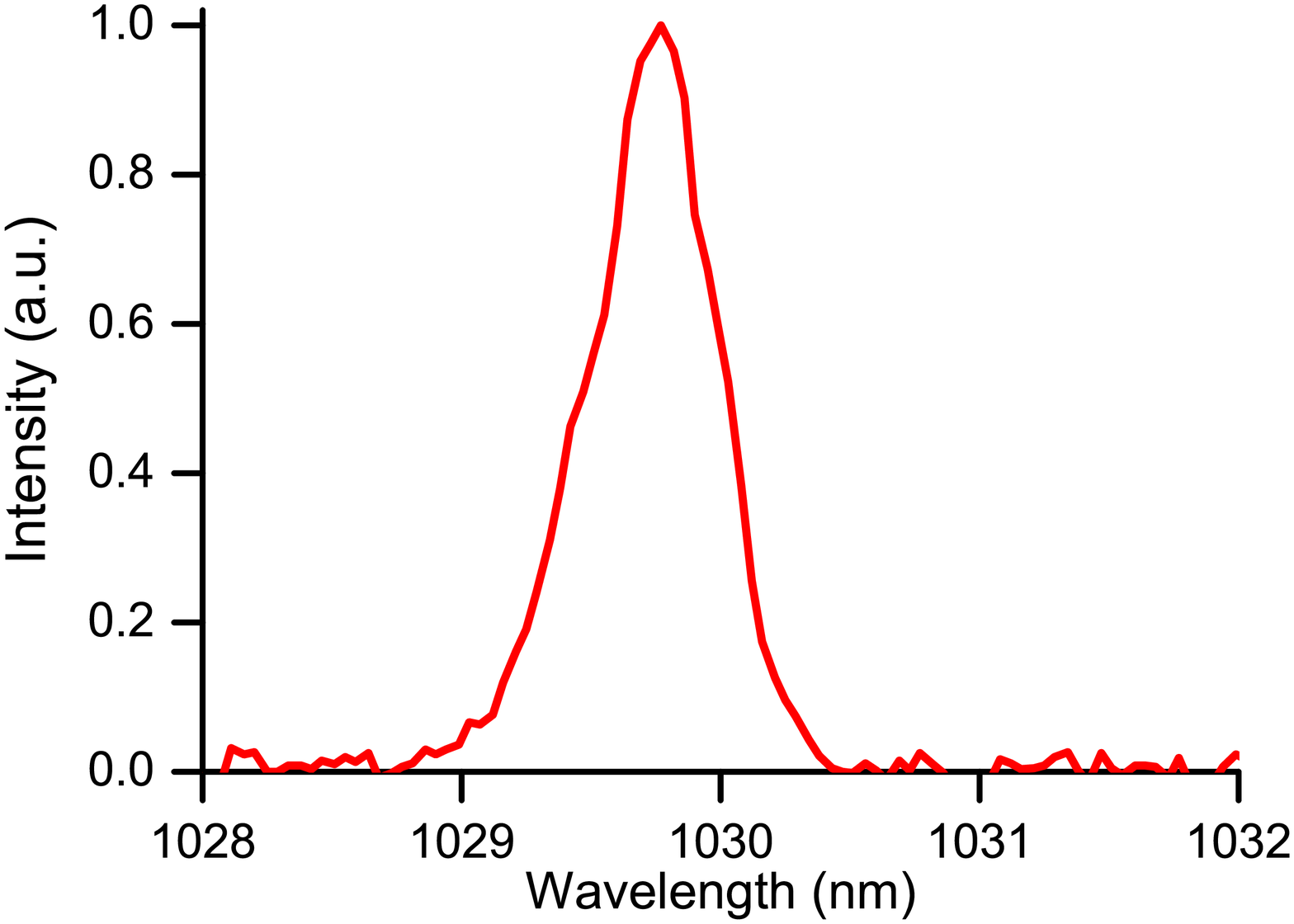}
                %\caption{Optical Spectrum}
                \label{fig:spectrum}
       			 }

      \subfigure[]{
                \includegraphics[width=0.4\textwidth]{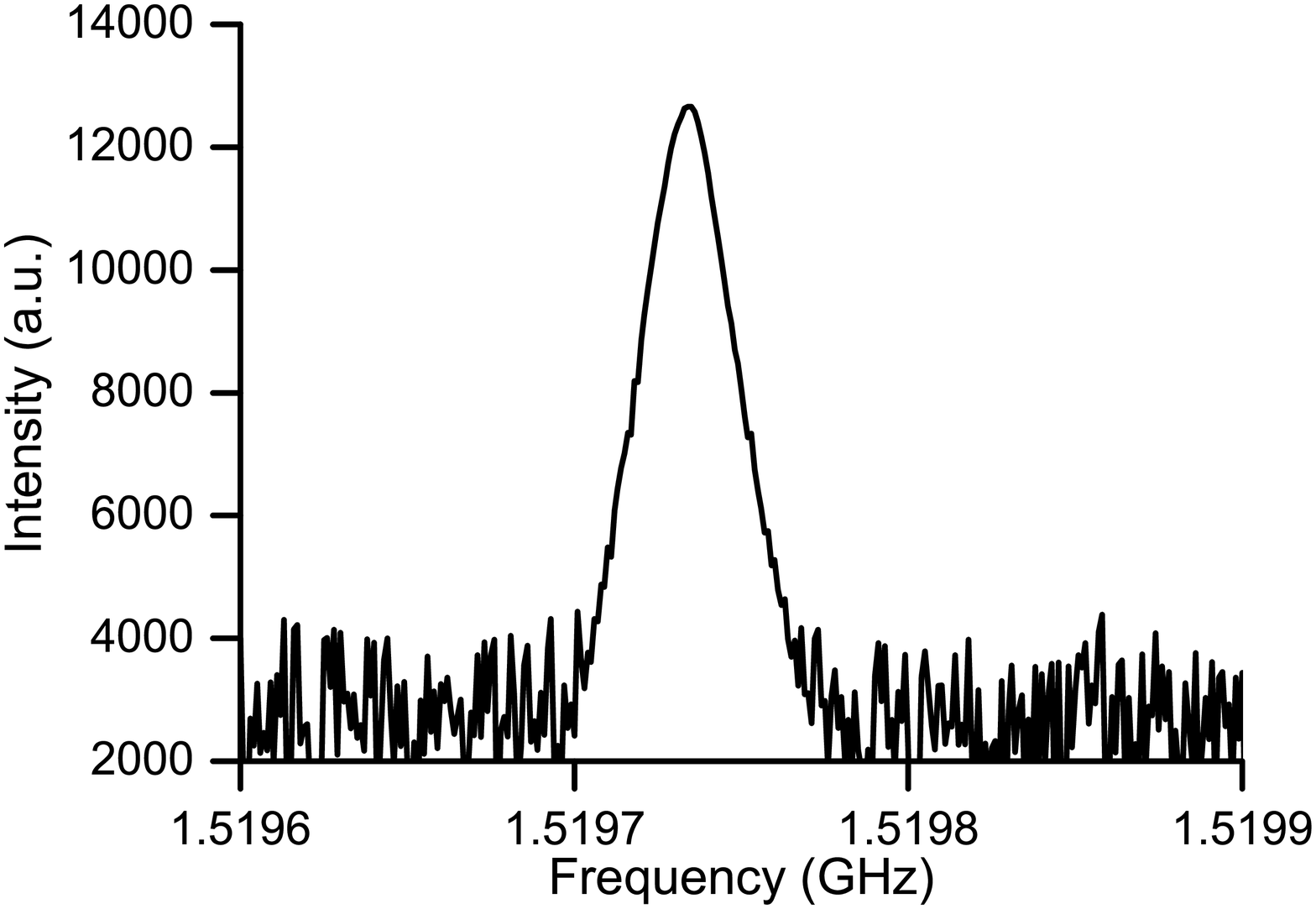}
                %\caption{RF Spectrum (1$^\text{st}$ Harmonic)}
                \label{fig:rf1}
     			 }
	\hspace*{\fill}
      \subfigure[]{
                \includegraphics[width=0.4\textwidth]{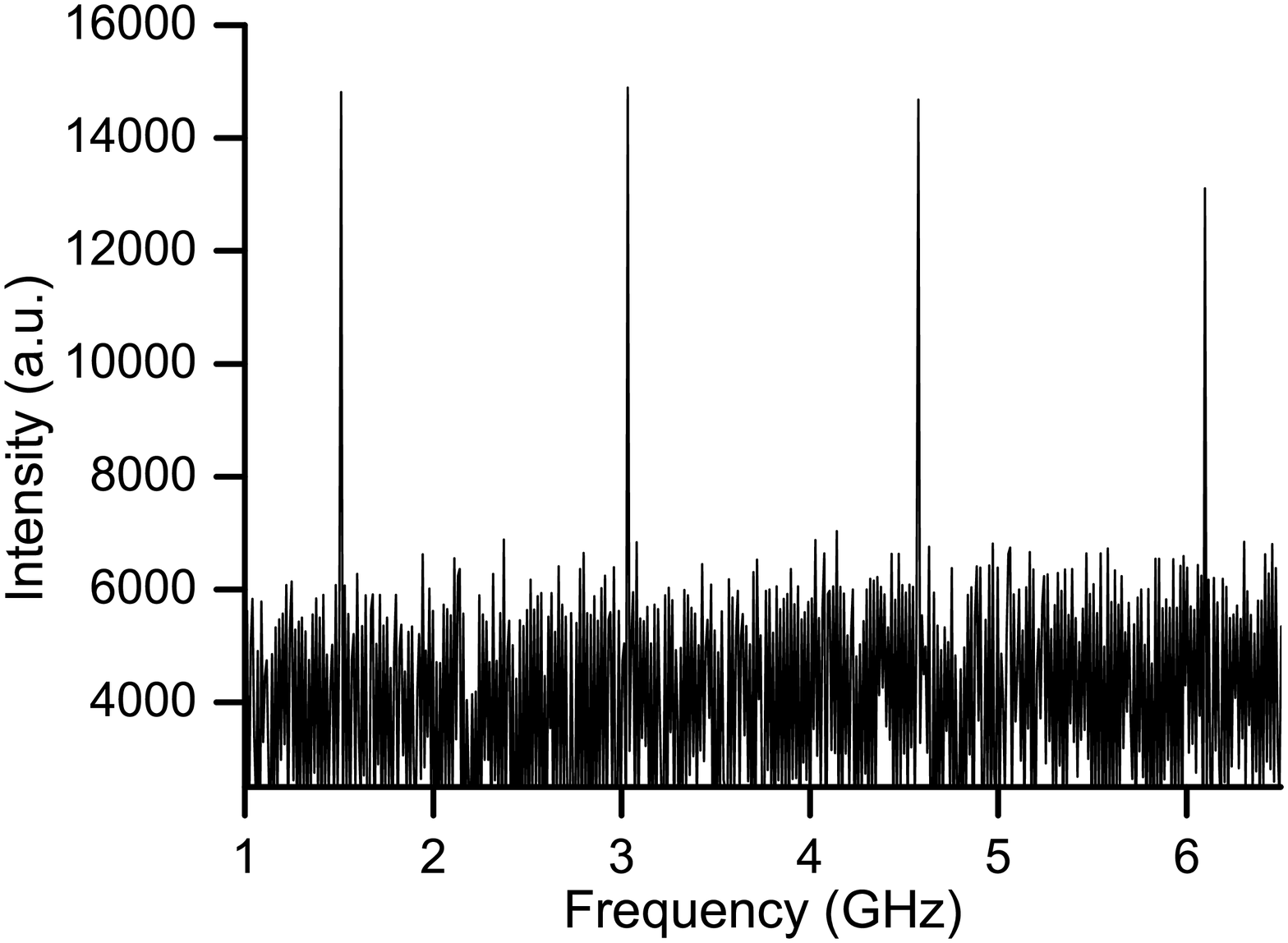}
                %\caption{RF Spectrum (1$^\text{st}$ -- 4$^\text{th}$ harmonics)}
                \label{fig:rf4}
			}      
  \caption{Characterisation of 4.5 ps modelocked VECSEL source emitting 1 W of average power at 1030 nm: (\ref{fig:autocorrelation}) Autocorrelation corresponding to a pulse duration of 4.5 ps. (\ref{fig:spectrum}) Optical spectrum corresponding to a centre wavelength of 1029.8 nm. (\ref{fig:rf1}) RF spectrum of the 1$^\text{st}$ harmonic corresponding to a repetition rate of 1.5 GHz. (\ref{fig:rf4}) RF spectrum of the first four harmonics.} 
        \label{fig:vecsel_data}
\end{figure}

Performance data are shown in Figure \ref{fig:vecsel_data}. The VECSEL emitted 4.5\,ps pulses at a repetition rate of 1.5 GHz, and was optimised for operation around 1030\,nm. The average output power was 1\,W corresponding to a peak power of 140\,W.  

\section{GHz repetition-rate photon-pair source}

\begin{figure}[h]
\centering
\includegraphics[width = 4.5cm]{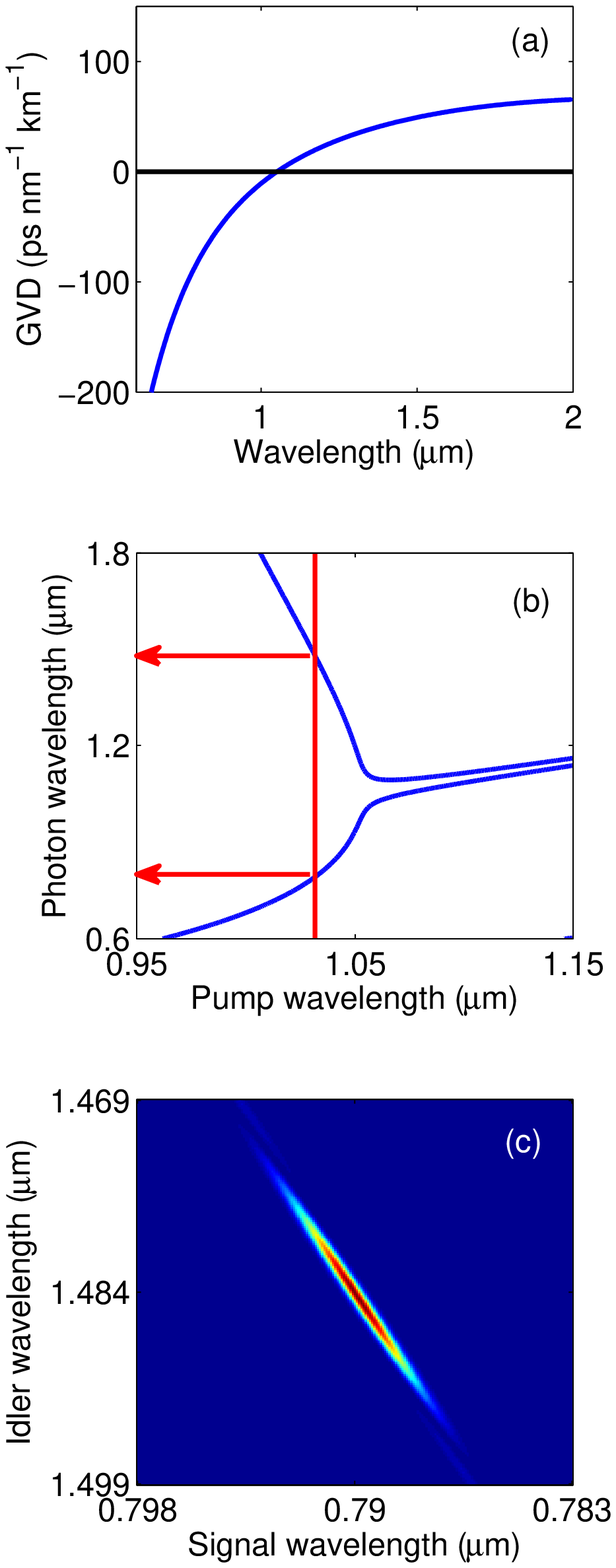}
\caption{(a) Modelled PCF group-velocity dispersion (GVD) and (b) corresponding FWM phasematching. (c) shows a calculated probability density plot for the resulting two-photon state when pumped at 1029\,nm.}
\label{fig:pcf}
\end{figure}

We have built a source of photon pairs based on the tunable VECSEL reported above and a PCF fabricated in Bath by the stack-and-draw technique. The PCF used in the experiments reported here was formed of a solid silica glass core surrounded by a cladding region formed by a triangular array of air holes in a silica glass matrix. The hole spacing (pitch) was approximately 3\,$\mu$m and the ratio of hole diameter to pitch was approximately 0.4. Its zero-dispersion wavelength was measured using white-light interferometry to be at 1058\,nm and a simulated dispersion profile based on these parameters is shown in Figure \ref{fig:pcf} \cite{Saitoh2005Empirical-relations-for-simple}. Using a PCF with bespoke dispersion allowed us to generate signal photons around 800\,nm where silicon detectors have high efficiency and idler photons within the low-loss window of conventional telecomms fibre around 1550\,nm.

\begin{figure}[h!]
\centering
\includegraphics[width = 12cm]{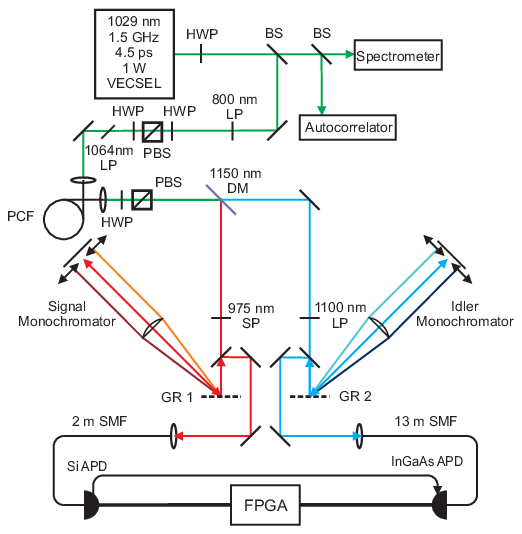}
\caption{VECSEL-pumped photon-pair source. See text for details.}
\label{fig:setup}
\end{figure}

The photon-pair source is shown in Figure \ref{fig:setup}. The beam from the VECSEL was cleaned by two long-wave-pass filters to remove any residual 808\,nm light originating in the pump diodes, and passed through two half-wave plates separated by a polarising beamsplitter for power and polarisation control. The VECSEL pulses were monitored continuously via pick-off beams directed to a spectrometer and an autocorrelator.

The pump pulses were coupled into and out of the PCF using aspheric lenses. Following the PCF, a half-wave plate and polariser selected only light co-polarised with the pump. The signal and idler were split with a dichroic mirror and the residual 1029\,nm pump was removed from both arms with additional interference filters. Subsequently both signal and idler passed through home-built grating monochromators to reduce any remaining background from the 808\,nm VECSEL pump diodes, 1029\,nm pulses, or spontaneous Raman scattering into the long-wavelength arm \cite{Fulconis2005High-brightness-single} and allow measurement of the signal and idler wavelengths. The photons were then coupled into conventional single-mode fibre (SMF-28 in the idler arm and SM800 in the signal arm).

Using CW diode lasers at 780\,nm and 1550\,nm we measured the total transmission from the output face of the PCF to the outputs of the single-mode fibres. We found over 90\% loss within the transmission bands of each arm, including all filtering and coupling. Approximately half of this loss originates in the grating monochromators, with the remaining half dominated by coupling into the single-mode fibres. The matching of the guided mode in the PCF to the two single-mode fibres was hampered by the chromatic variation in focal length of the aspheric lens after the PCF; as a result it was necessary to compromise between good coupling in the long- and short-wavelength arms. Loss from the other components following the PCF was negligible.

% numbers for idler arm?

The 800\,nm photons were routed through 2\,m of SM800 to a free-running silicon avalanche photodiode (Perkin-Elmer SPCM). An InGaAs detector (idQuantique id201) was used to monitor the 1550\,nm arm. It could not be operated in free-running mode and had a maximum gate frequency of 1\,MHz -- over three orders of magnitude lower than the repetition frequency of the pump laser. Therefore the output pulses from the Si APD were used to gate the InGaAs APD. The 1550\,nm photons were delayed in 13\,m of fibre to provide sufficient time for the InGaAs detector to become active before the corresponding photon arrived. The count rates from both detectors were monitored by a home-built field-programmable gate array (FPGA) coincidence counting unit.

\begin{figure}[h]
	\subfigure[]{
	\includegraphics[width = 0.4\textwidth]{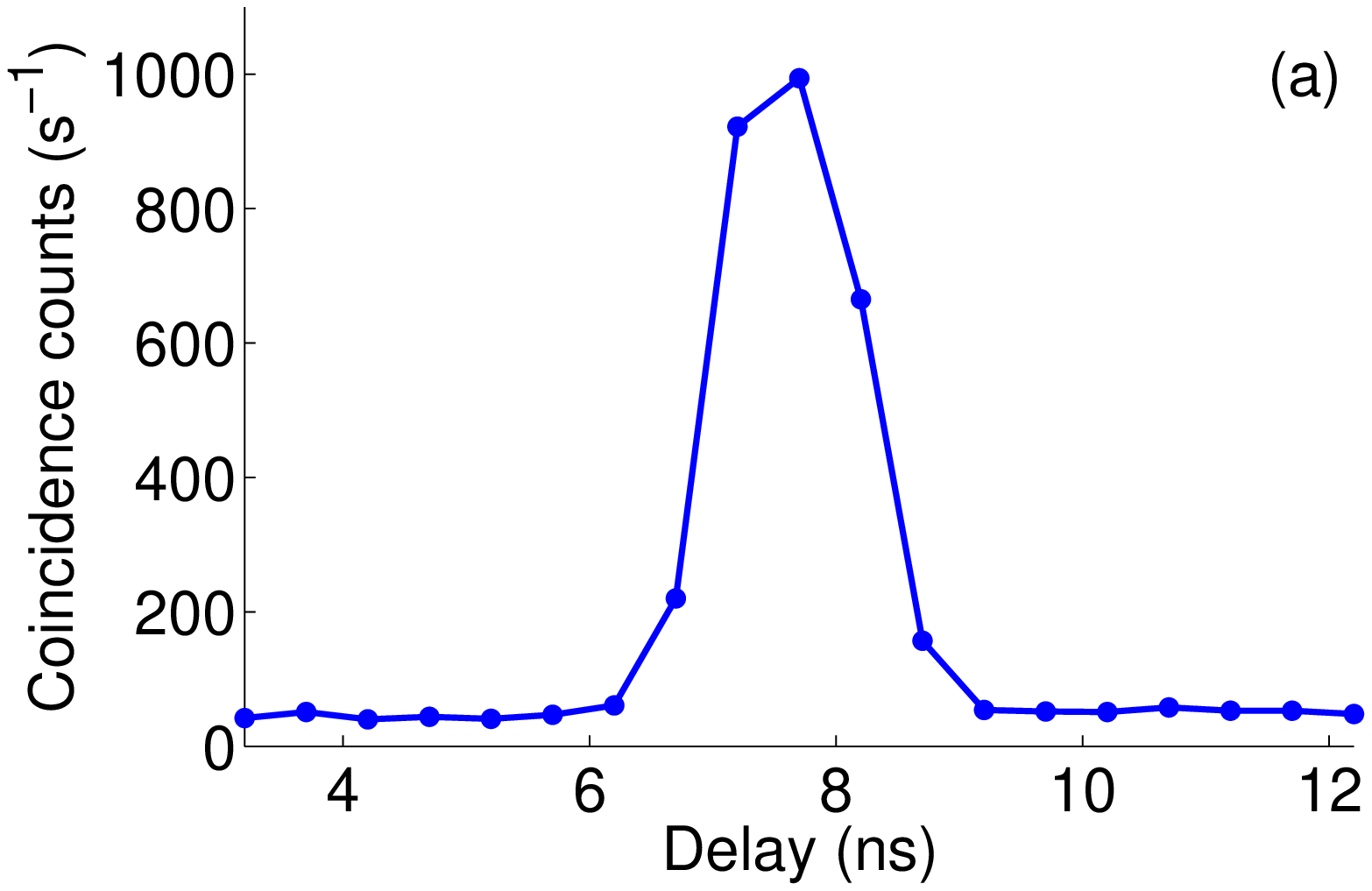}
			}
	\hspace*{\fill}
	\subfigure[]{
	\includegraphics[width = 0.4\textwidth]{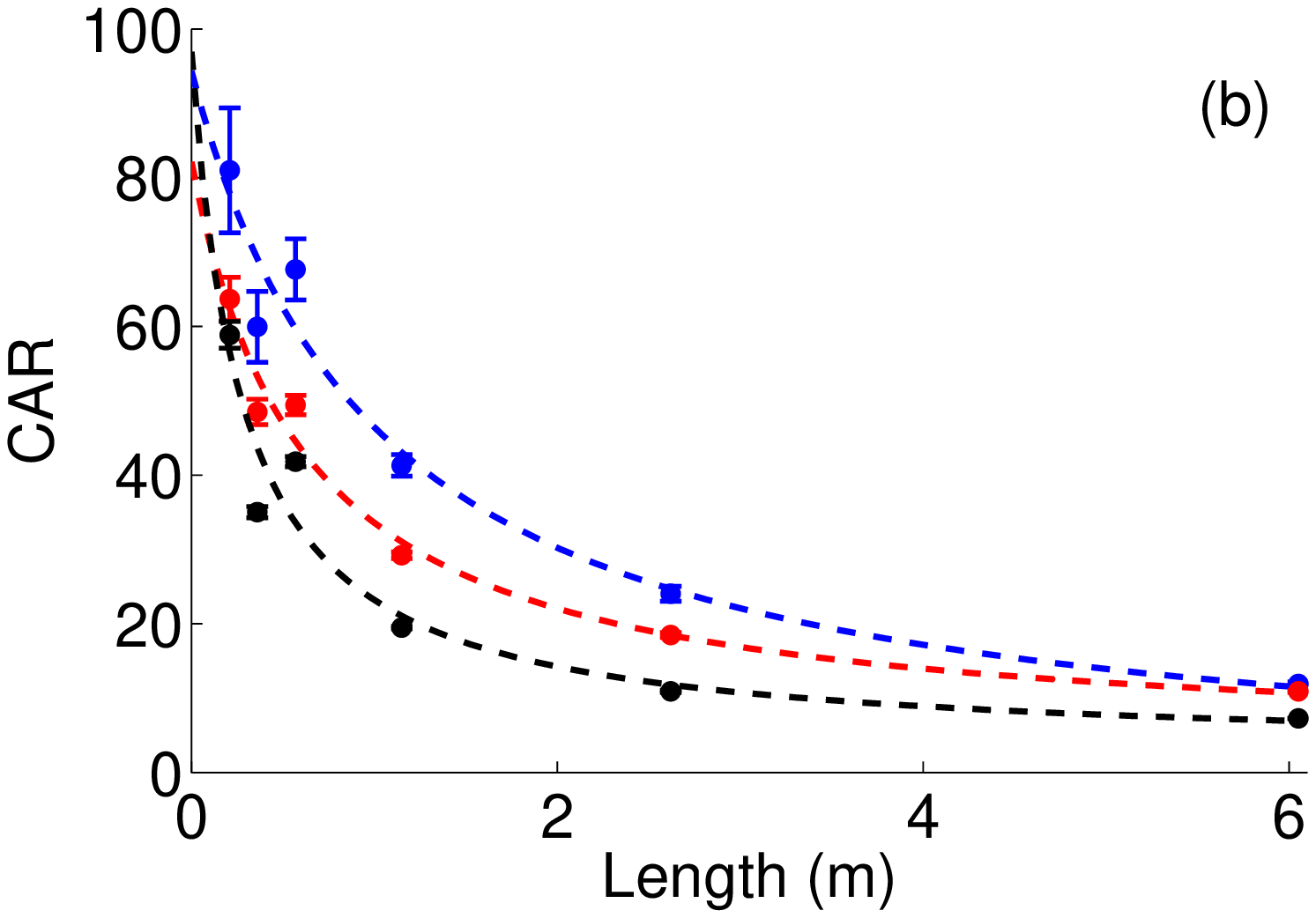}
			}
\caption{(a) A typical plot of InGaAs APD count rates (equal to the coincidence count rate) against internal detector delay. (b) Coincidence-to-accidental ratio as a function of PCF length for average pump powers of 50\,mW (blue), 100\,mW (red), and 150\,mW (black). Dotted lines are fits proportional to $1/L$.}
\label{fig:source_data}
\end{figure}

Figure \ref{fig:source_data} shows a typical plot of the InGaAs detector count rate as a function of the internal delay between the arrival of the trigger pulse and the detection window opening. With the delay set to 7.5\,ns, the detection window of the InGaAs APD encompassed the arrival times of photons generated by the same pump pulse as those detected at the Si APD. The height of this peak therefore showed the number of coincidence counts per second between the signal and idler arms. Away from this peak is a constant background level of accidental coincidence events arising from uncorrelated generation and detection processes. The ratio of coincidence counts to accidental background events is known as the coincidence-to-accidentals ratio or CAR and is a strong indicator of source performance.

The gate width of the InGaAs APD was set to its minimum value of 2.5\,ns. Because the detection events at the Si APD were used to trigger the InGaAs detector, any photons arriving within the gate of the InGaAs detector were registered as coincidence counts and the width of the coincidence peak was set by the detector gate width. It can be seen from Figure \ref{fig:source_data} that the resulting effective coincidence window is less than 2\,ns, or approximately three times the laser repetition period. Therefore for each detection event at the Si APD there were three possible pump pulses within which any corresponding photons counted by the InGaAs APD could have originated. Nevertheless, the high-contrast coincidence signal at 7.5\,ns and constant background at all other delays is strong evidence of a correlated generation process from a single pump pulse as one would expect from FWM. Within the limitations of the detectors available, there was no possibility of reducing the coincidence window to a single pump pulse, however suitable avalanche photodiodes have been gated at over 2\,GHz \cite{Zhang20102.23-GHz-gating-InGaAs/InP}.

With the pump wavelength set to 1029\,nm, the central wavelengths and bandwidths of the monochromators were adjusted to maximise the CAR. The central wavelengths of signal and idler were found to be approximately 780\,nm and 1515\,nm respectively.

In order to optimise the length of the PCF in the photon-pair source, we began with a 6\,m length of PCF and cut it back while monitoring the CAR. The results for three different power levels are shown in Figure \ref{fig:source_data}. It can be seen that the CAR was reasonably high at all times, never dropping below 7. The CAR rose rapidly as the PCF length was reduced below 1\,m indicating a significant improvement in source performance; the highest measured value of the CAR was over 80, using 50\,mW average pump power and a PCF length of 210\,mm, and at an absolute coincidence count rate of almost 1000 pairs/s. This fibre length corresponded to approximately half the calculated walk-off length between the signal and idler and the 4.5\,ps pump pulses (450\,mm and 340\,mm respectively).

The absolute count rates were limited by the high level of loss following the PCF, particularly in the idler arm. Nevertheless the highest recorded coincidence count rate was 9600 pairs/s with a CAR of 19 from 150\,mW average pump power in a fibre length of  1.15\,m. Subtracting the measured accidental coincidence rate (480/s) and taking into account the loss after the PCF of over 90\% in each arm and detection efficiencies of 60\% (signal) and 25\% (idler) implies a pair generation rate of at least $6\times10^6$\,pairs/s and a generated brightness of over $4\times10^4$\,pairs/s/mW within bandwidths of approximately 4\,nm in the signal arm and 6\,nm in the idler.

 In future, we expect to be able to reduce the loss and significantly increase the absolute count rates by moving to an all-fibre spliced system, thereby extending the source performance far beyond this proof-of-principle demonstration.

\section{Additional capabilities -- tuning and state engineering}

With the PCF reported above, tuning the VECSEL over its full range of $1027\pm4.5$\,nm would allow generation of photon pairs from 751 -- 790\,nm (1600 -- 1484\,nm) for the signal (idler). More significantly than this, the tunability and bandwidth of the VECSEL can be exploited for for quantum state engineering -- controlling the properties of the photon pairs and heralded single photons through modification of the fibre dispersion \cite{Garay-Palmett2007Photon-pair-state-preparation}. The one-pair component of the state that results from FWM can be written
\begin{equation}
\ket{1_\text{s}, 1_\text{i}} = \ket{\psi(\omega_\text{s}, \omega_\text{i})} = \int_{-\infty}^{\infty} d\omega_\text{s} d\omega_\text{i} f(\omega_\text{s}, \omega_\text{i}) \ket{\omega_\text{s}, \omega_\text{i}},
\end{equation}
where $f(\omega_\text{s}, \omega_\text{i})$ is the two-photon amplitude or joint spectrum. In general this amplitude contains correlation between the signal and idler in both the frequency and time domains. The correlation arises because the photons are created together in a parametric process subject to energy and momentum conservation. As a result, detecting each photon yields distinguishing information about its twin, projecting the remaining heralded single photons into mixed quantum states that are largely useless for information-processing tasks.

In order to herald single photons in pure quantum states the photon-pair amplitude must be uncorrelated \cite{Grice2001Eliminating-frequency-and-space-time}. Producing an uncorrelated two-photon state is not trivial, and there are two critical requirements: control of the material dispersion and sufficient pump bandwidth to ensure that energy conservation does not dominate the joint state \cite{Mosley2008Conditional-preparation-of-single, Cohen2009Tailored-Photon-Pair-Generation, Soller2011High-performance-single-photon-generation, Cui2012Minimizing-the-frequency-correlation, Yabuno2012Four-photon-quantum-interferometry}. The VECSEL has enough bandwidth, and, while we can fabricate PCF with a wide range of different dispersion profiles, fine control of the dispersion is a major challenge. Even a small degree of of tunability in the pump laser is hugely beneficial to compensate for small errors in fabrication and allow the precise phasematching requirements to be accessed.

\section{Conclusion}

FWM in PCF is an attractive method for producing photon pairs for quantum-information applications, as it combines the simplicity of nonlinear wave-mixing techniques with guided-wave architecture to allow high-efficiency pair generation. We have demonstrated a new combination of ultra-high repetition-rate VECSEL and PCF that provides new possibilities in reducing detrimental multiple-pair emission while maintaining good count rates and a very high CAR. The small footprint and simplicity of the laser are attractive for integration while its bandwidth and tunability, in conjunction with the flexible dispersion offered by PCF, make it amenable to photonic quantum-state engineering.

\section{Acknowledgements}

We thank William Wadsworth for providing the PCF and Wolfgang Stolz, Bernardette Kunert and Bernd Heinen from NAsP III/V for growth and preparation of the VECSEL gain chip. This work was supported by the EPSRC First Grant scheme (EP/K022407/1). O.J.M. acknowledges an EPSRC studentship and K.G.W. acknowledges EPSRC for his personal fellowship (EP/J017043/1).

%\bibliographystyle{elsarticle/elsarticle-num}
%\bibliography{../../../../Papers/Bibliography/general}

\begin{thebibliography}{10}
\expandafter\ifx\csname url\endcsname\relax
  \def\url#1{\texttt{#1}}\fi
\expandafter\ifx\csname urlprefix\endcsname\relax\def\urlprefix{URL }\fi
\expandafter\ifx\csname href\endcsname\relax
  \def\href#1#2{#2} \def\path#1{#1}\fi

\bibitem{OBrien2009Photonic-quantum-technologies}
J.~L. O'Brien, A.~Furusawa, J.~Vuckovic,
  \href{http://dx.doi.org/10.1038/nphoton.2009.229}{Photonic quantum
  technologies}, Nat Photon 3~(12) (2009) 687--695.
\newline\urlprefix\url{http://dx.doi.org/10.1038/nphoton.2009.229}

\bibitem{Eisaman2011Single-photon-sources-and-detectors}
M.~D. Eisaman, J.~Fan, A.~Migdall, S.~V. Polyakov,
  \href{http://link.aip.org/link/?RSI/82/071101/1}{Single-photon sources and
  detectors}, Review of Scientific Instruments 82~(7) (2011) 071101.
\newblock \href {http://dx.doi.org/10.1063/1.3610677}
  {\path{doi:10.1063/1.3610677}}.
\newline\urlprefix\url{http://link.aip.org/link/?RSI/82/071101/1}

\bibitem{Pan2012Multiphoton-entanglement-and-interferometry}
J.-W. Pan, Z.-B. Chen, C.-Y. Lu, H.~Weinfurter, A.~Zeilinger,
  M.~\ifmmode~\dot{Z}\else \.{Z}\fi{}ukowski,
  \href{http://link.aps.org/doi/10.1103/RevModPhys.84.777}{Multiphoton
  entanglement and interferometry}, Rev. Mod. Phys. 84 (2012) 777--838.
\newblock \href {http://dx.doi.org/10.1103/RevModPhys.84.777}
  {\path{doi:10.1103/RevModPhys.84.777}}.
\newline\urlprefix\url{http://link.aps.org/doi/10.1103/RevModPhys.84.777}

\bibitem{Soller2010Bridging-visible-and-telecom}
C.~S\"oller, B.~Brecht, P.~J. Mosley, L.~Y. Zang, A.~Podlipensky, N.~Y. Joly,
  P.~S.~J. Russell, C.~Silberhorn, Bridging visible and telecom wavelengths
  with a single-mode broadband photon pair source, Phys. Rev. A 81~(3) (2010)
  031801.
\newblock \href {http://dx.doi.org/10.1103/PhysRevA.81.031801}
  {\path{doi:10.1103/PhysRevA.81.031801}}.

\bibitem{Xiong2011Slow-light-Enhanced-Correlated}
C.~{Xiong}, C.~{Monat}, A.~S. {Clark}, C.~{Grillet}, G.~D. {Marshall}, M.~J.
  {Steel}, J.~{Li}, L.~{O'Faolain}, T.~F. {Krauss}, J.~G. {Rarity}, B.~J.
  {Eggleton}, {Slow-light Enhanced Correlated Photon-Pair Generation in
  Silicon}, ArXiv e-prints\href {http://arxiv.org/abs/1106.3593}
  {\path{arXiv:1106.3593}}.

\bibitem{Collins2012Low-Raman-noise-correlated-photon-pair}
M.~J. Collins, A.~S. Clark, J.~He, D.-Y. Choi, R.~J. Williams, A.~C. Judge,
  S.~J. Madden, M.~J. Withford, M.~J. Steel, B.~Luther-Davies, C.~Xiong, B.~J.
  Eggleton, \href{http://ol.osa.org/abstract.cfm?URI=ol-37-16-3393}{Low
  raman-noise correlated photon-pair generation in a dispersion-engineered
  chalcogenide as2s3 planar waveguide}, Opt. Lett. 37~(16) (2012) 3393--3395.
\newblock \href {http://dx.doi.org/10.1364/OL.37.003393}
  {\path{doi:10.1364/OL.37.003393}}.
\newline\urlprefix\url{http://ol.osa.org/abstract.cfm?URI=ol-37-16-3393}

\bibitem{Steinlechner2012A-high-brightness-source-of-polarization-entangled}
F.~Steinlechner, P.~Trojek, M.~Jofre, H.~Weier, D.~Perez, T.~Jennewein,
  R.~Ursin, J.~Rarity, M.~W. Mitchell, J.~P. Torres, H.~Weinfurter, V.~Pruneri,
  \href{http://www.opticsexpress.org/abstract.cfm?URI=oe-20-9-9640}{A
  high-brightness source of polarization-entangled photons optimized for
  applications in free space}, Opt. Express 20~(9) (2012) 9640--9649.
\newblock \href {http://dx.doi.org/10.1364/OE.20.009640}
  {\path{doi:10.1364/OE.20.009640}}.
\newline\urlprefix\url{http://www.opticsexpress.org/abstract.cfm?URI=oe-20-9-9640}

\bibitem{Karpinski2012Dispersion-based-control-of-modal}
M.~Karpi\'{n}ski, C.~Radzewicz, K.~Banaszek,
  \href{http://ol.osa.org/abstract.cfm?URI=ol-37-5-878}{Dispersion-based
  control of modal characteristics for parametric down-conversion in a
  multimode waveguide}, Opt. Lett. 37~(5) (2012) 878--880.
\newblock \href {http://dx.doi.org/10.1364/OL.37.000878}
  {\path{doi:10.1364/OL.37.000878}}.
\newline\urlprefix\url{http://ol.osa.org/abstract.cfm?URI=ol-37-5-878}

\bibitem{Pomarico2012MHz-rate-and-efficient-synchronous}
E.~Pomarico, B.~Sanguinetti, T.~Guerreiro, R.~Thew, H.~Zbinden,
  \href{http://www.opticsexpress.org/abstract.cfm?URI=oe-20-21-23846}{Mhz rate
  and efficient synchronous heralding of single photons at telecom
  wavelengths}, Opt. Express 20~(21) (2012) 23846--23855.
\newblock \href {http://dx.doi.org/10.1364/OE.20.023846}
  {\path{doi:10.1364/OE.20.023846}}.
\newline\urlprefix\url{http://www.opticsexpress.org/abstract.cfm?URI=oe-20-21-23846}

\bibitem{McMillan2013Two-photon-interference-between}
A.~R. McMillan, L.~Labont, A.~S. Clark, B.~Bell, O.~Alibart, A.~Martin, W.~J.
  Wadsworth, S.~Tanzilli, J.~G. Rarity,
  \href{http://dx.doi.org/10.1038/srep02032}{Two-photon interference between
  disparate sources for quantum networking}, Sci. Rep. 3.
\newline\urlprefix\url{http://dx.doi.org/10.1038/srep02032}

\bibitem{Mosley2008Heralded-Generation-of-Ultrafast}
P.~J. Mosley, J.~S. Lundeen, B.~J. Smith, P.~Wasylczyk, A.~B. U'Ren,
  C.~Silberhorn, I.~A. Walmsley,
  \href{http://link.aps.org/abstract/PRL/v100/e133601}{Heralded generation of
  ultrafast single photons in pure quantum states}, Physical Review Letters
  100~(13) (2008) 133601.
\newblock \href {http://dx.doi.org/10.1103/PhysRevLett.100.133601}
  {\path{doi:10.1103/PhysRevLett.100.133601}}.
\newline\urlprefix\url{http://link.aps.org/abstract/PRL/v100/e133601}

\bibitem{Ling2009Mode-expansion-and-Bragg}
A.~Ling, J.~Chen, J.~Fan, A.~Migdall,
  \href{http://www.opticsexpress.org/abstract.cfm?URI=oe-17-23-21302}{Mode
  expansion and bragg filtering for a high-fidelity fiber-based photon-pair
  source}, Opt. Express 17~(23) (2009) 21302--21312.
\newblock \href {http://dx.doi.org/10.1364/OE.17.021302}
  {\path{doi:10.1364/OE.17.021302}}.
\newline\urlprefix\url{http://www.opticsexpress.org/abstract.cfm?URI=oe-17-23-21302}

\bibitem{Eckstein2011Highly-Efficient-Single-Pass}
A.~Eckstein, A.~Christ, P.~J. Mosley, C.~Silberhorn, Highly efficient
  single-pass source of pulsed single-mode twin beams of light, Phys. Rev.
  Lett. 106~(1) (2011) 013603.
\newblock \href {http://dx.doi.org/10.1103/PhysRevLett.106.013603}
  {\path{doi:10.1103/PhysRevLett.106.013603}}.

\bibitem{Clark2011Intrinsically-narrowband-pair}
A.~Clark, B.~Bell, J.~Fulconis, M.~M. Halder, B.~Cemlyn, O.~Alibart, C.~Xiong,
  W.~J. Wadsworth, J.~G. Rarity,
  \href{http://stacks.iop.org/1367-2630/13/i=6/a=065009}{Intrinsically
  narrowband pair photon generation in microstructured fibres}, New Journal of
  Physics 13~(6) (2011) 065009.
\newline\urlprefix\url{http://stacks.iop.org/1367-2630/13/i=6/a=065009}

\bibitem{Tanida2012Highly-indistinguishable-heralded}
M.~Tanida, R.~Okamoto, S.~Takeuchi,
  \href{http://www.opticsexpress.org/abstract.cfm?URI=oe-20-14-15275}{Highly
  indistinguishable heralded single-photon sources using parametric down
  conversion}, Opt. Express 20~(14) (2012) 15275--15285.
\newblock \href {http://dx.doi.org/10.1364/OE.20.015275}
  {\path{doi:10.1364/OE.20.015275}}.
\newline\urlprefix\url{http://www.opticsexpress.org/abstract.cfm?URI=oe-20-14-15275}

\bibitem{Jin2013Widely-tunable-single}
R.-B. Jin, R.~Shimizu, K.~Wakui, H.~Benichi, M.~Sasaki,
  \href{http://www.opticsexpress.org/abstract.cfm?URI=oe-21-9-10659}{Widely
  tunable single photon source with high purity at telecom wavelength}, Opt.
  Express 21~(9) (2013) 10659--10666.
\newblock \href {http://dx.doi.org/10.1364/OE.21.010659}
  {\path{doi:10.1364/OE.21.010659}}.
\newline\urlprefix\url{http://www.opticsexpress.org/abstract.cfm?URI=oe-21-9-10659}

\bibitem{Christ2012Limits-on-the-deterministic-creation}
A.~Christ, C.~Silberhorn,
  \href{http://link.aps.org/doi/10.1103/PhysRevA.85.023829}{Limits on the
  deterministic creation of pure single-photon states using parametric
  down-conversion}, Phys. Rev. A 85 (2012) 023829.
\newblock \href {http://dx.doi.org/10.1103/PhysRevA.85.023829}
  {\path{doi:10.1103/PhysRevA.85.023829}}.
\newline\urlprefix\url{http://link.aps.org/doi/10.1103/PhysRevA.85.023829}

\bibitem{Morris2012A-wavelength-tunable-2-ps}
O.~J. Morris, K.~G. Wilcox, C.~R. Head, A.~P. Turnbull, P.~J. Mosley, A.~H.
  Quarterman, H.~J. Kbashi, I.~Farrer, H.~E. Beere, D.~A. Ritchie, A.~C.
  Tropper, \href{http://dx.doi.org/10.1117/12.908337}{A wavelength tunable 2-ps
  pulse vecsel} (2012).
\newblock \href {http://dx.doi.org/10.1117/12.908337}
  {\path{doi:10.1117/12.908337}}.
\newline\urlprefix\url{http://dx.doi.org/10.1117/12.908337}

\bibitem{Quarterman2009A-passively-mode-locked-external-cavity}
A.~H. Quarterman, K.~G. Wilcox, V.~Apostolopoulos, Z.~Mihoubi, S.~P. Elsmere,
  I.~Farrer, D.~A. Ritchie, A.~Tropper,
  \href{http://dx.doi.org/10.1038/nphoton.2009.216}{A passively mode-locked
  external-cavity semiconductor laser emitting 60-fs pulses}, Nat Photon 3~(12)
  (2009) 729--731.
\newline\urlprefix\url{http://dx.doi.org/10.1038/nphoton.2009.216}

\bibitem{Wilcox20134.35-kW-peak-power}
K.~G. Wilcox, A.~C. Tropper, H.~E. Beere, D.~A. Ritchie, B.~Kunert, B.~Heinen,
  W.~Stolz,
  \href{http://www.opticsexpress.org/abstract.cfm?URI=oe-21-2-1599}{4.35 kw
  peak power femtosecond pulse mode-locked vecsel for supercontinuum
  generation}, Opt. Express 21~(2) (2013) 1599--1605.
\newblock \href {http://dx.doi.org/10.1364/OE.21.001599}
  {\path{doi:10.1364/OE.21.001599}}.
\newline\urlprefix\url{http://www.opticsexpress.org/abstract.cfm?URI=oe-21-2-1599}

\bibitem{Wilcox2011Repetition-frequency-tunable-mode-locked-surface}
K.~G. Wilcox, A.~H. Quarterman, H.~E. Beere, D.~A. Ritchie, A.~C. Tropper,
  \href{http://www.opticsexpress.org/abstract.cfm?URI=oe-19-23-23453}{Repetition-frequency-tunable
  mode-locked surface emitting semiconductor laser between 2.78 and 7.87 ghz},
  Opt. Express 19~(23) (2011) 23453--23459.
\newblock \href {http://dx.doi.org/10.1364/OE.19.023453}
  {\path{doi:10.1364/OE.19.023453}}.
\newline\urlprefix\url{http://www.opticsexpress.org/abstract.cfm?URI=oe-19-23-23453}

\bibitem{Sansoni2012Two-Particle-Bosonic-Fermionic-Quantum}
L.~Sansoni, F.~Sciarrino, G.~Vallone, P.~Mataloni, A.~Crespi, R.~Ramponi,
  R.~Osellame,
  \href{http://link.aps.org/doi/10.1103/PhysRevLett.108.010502}{Two-particle
  bosonic-fermionic quantum walk via integrated photonics}, Phys. Rev. Lett.
  108 (2012) 010502.
\newblock \href {http://dx.doi.org/10.1103/PhysRevLett.108.010502}
  {\path{doi:10.1103/PhysRevLett.108.010502}}.
\newline\urlprefix\url{http://link.aps.org/doi/10.1103/PhysRevLett.108.010502}

\bibitem{Spring2013Boson-Sampling-on-a-Photonic}
J.~B. Spring, B.~J. Metcalf, P.~C. Humphreys, W.~S. Kolthammer, X.-M. Jin,
  M.~Barbieri, A.~Datta, N.~Thomas-Peter, N.~K. Langford, D.~Kundys, J.~C.
  Gates, B.~J. Smith, P.~G.~R. Smith, I.~A. Walmsley,
  \href{http://www.sciencemag.org/content/339/6121/798.abstract}{Boson sampling
  on a photonic chip}, Science 339~(6121) (2013) 798--801.
\newblock \href
  {http://arxiv.org/abs/http://www.sciencemag.org/content/339/6121/798.full.pdf}
  {\path{arXiv:http://www.sciencemag.org/content/339/6121/798.full.pdf}}, \href
  {http://dx.doi.org/10.1126/science.1231692}
  {\path{doi:10.1126/science.1231692}}.
\newline\urlprefix\url{http://www.sciencemag.org/content/339/6121/798.abstract}

\bibitem{Broome2013Photonic-Boson-Sampling}
M.~A. Broome, A.~Fedrizzi, S.~Rahimi-Keshari, J.~Dove, S.~Aaronson, T.~C.
  Ralph, A.~G. White,
  \href{http://www.sciencemag.org/content/339/6121/794.abstract}{Photonic boson
  sampling in a tunable circuit}, Science 339~(6121) (2013) 794--798.
\newblock \href
  {http://arxiv.org/abs/http://www.sciencemag.org/content/339/6121/794.full.pdf}
  {\path{arXiv:http://www.sciencemag.org/content/339/6121/794.full.pdf}}, \href
  {http://dx.doi.org/10.1126/science.1231440}
  {\path{doi:10.1126/science.1231440}}.
\newline\urlprefix\url{http://www.sciencemag.org/content/339/6121/794.abstract}

\bibitem{Chen2005Two-photon-state-generation-via-four-wave}
J.~Chen, X.~Li, P.~Kumar,
  \href{http://link.aps.org/doi/10.1103/PhysRevA.72.033801}{Two-photon-state
  generation via four-wave mixing in optical fibers}, Phys. Rev. A 72 (2005)
  033801.
\newblock \href {http://dx.doi.org/10.1103/PhysRevA.72.033801}
  {\path{doi:10.1103/PhysRevA.72.033801}}.
\newline\urlprefix\url{http://link.aps.org/doi/10.1103/PhysRevA.72.033801}

\bibitem{Lin2007Photon-pair-generation-in-optical}
Q.~Lin, F.~Yaman, G.~P. Agrawal,
  \href{http://link.aps.org/abstract/PRA/v75/e023803}{Photon-pair generation in
  optical fibers through four-wave mixing: Role of raman scattering and pump
  polarization}, Physical Review A 75~(2) (2007) 023803.
\newblock \href {http://dx.doi.org/10.1103/PhysRevA.75.023803}
  {\path{doi:10.1103/PhysRevA.75.023803}}.
\newline\urlprefix\url{http://link.aps.org/abstract/PRA/v75/e023803}

\bibitem{Migdall2002Tailoring-single-photon-and-multiphoton}
A.~L. Migdall, D.~Branning, S.~Castelletto,
  \href{http://link.aps.org/doi/10.1103/PhysRevA.66.053805}{Tailoring
  single-photon and multiphoton probabilities of a single-photon on-demand
  source}, Phys. Rev. A 66 (2002) 053805.
\newblock \href {http://dx.doi.org/10.1103/PhysRevA.66.053805}
  {\path{doi:10.1103/PhysRevA.66.053805}}.
\newline\urlprefix\url{http://link.aps.org/doi/10.1103/PhysRevA.66.053805}

\bibitem{Broome2011Reducing-multi-photon-rates}
M.~A. Broome, M.~P. Almeida, A.~Fedrizzi, A.~G. White,
  \href{http://www.opticsexpress.org/abstract.cfm?URI=oe-19-23-22698}{Reducing
  multi-photon rates in pulsed down-conversion by temporal multiplexing}, Opt.
  Express 19~(23) (2011) 22698--22708.
\newblock \href {http://dx.doi.org/10.1364/OE.19.022698}
  {\path{doi:10.1364/OE.19.022698}}.
\newline\urlprefix\url{http://www.opticsexpress.org/abstract.cfm?URI=oe-19-23-22698}

\bibitem{Saitoh2005Empirical-relations-for-simple}
K.~Saitoh, M.~Koshiba,
  \href{http://www.opticsexpress.org/abstract.cfm?URI=oe-13-1-267}{Empirical
  relations for simple design of photonic crystal fibers}, Opt. Express 13~(1)
  (2005) 267--274.
\newline\urlprefix\url{http://www.opticsexpress.org/abstract.cfm?URI=oe-13-1-267}

\bibitem{Fulconis2005High-brightness-single}
J. Fulconis, O. Alibart, W. Wadsworth, P. Russell, J. Rarity,
\href{}{High brightness single mode source of correlated photon pairs using a photonic crystal fiber}, Opt. Express 13 (2005) 7572--7582.

\bibitem{Zhang20102.23-GHz-gating-InGaAs/InP}
J.~Zhang, P.~Eraerds, N.~Walenta, C.~Barreiro, R.~Thew, H.~Zbinden,
  \href{http://dx.doi.org/10.1117/12.862118}{2.23 ghz gating ingaas/inp
  single-photon avalanche diode for quantum key distribution} (2010).
\newblock \href {http://dx.doi.org/10.1117/12.862118}
  {\path{doi:10.1117/12.862118}}.
\newline\urlprefix\url{http://dx.doi.org/10.1117/12.862118}

\bibitem{Garay-Palmett2007Photon-pair-state-preparation}
K.~Garay-Palmett, H.~J. McGuinness, O.~Cohen, J.~S. Lundeen, R.~Rangel-Rojo,
  A.~B. U'ren, M.~G. Raymer, C.~J. McKinstrie, S.~Radic, I.~A. Walmsley,
  \href{http://www.opticsexpress.org/abstract.cfm?URI=oe-15-22-14870}{Photon
  pair-state preparation with tailored spectral properties by spontaneous
  four-wave mixing in photonic-crystal fiber}, Opt. Express 15~(22) (2007)
  14870--14886.
\newline\urlprefix\url{http://www.opticsexpress.org/abstract.cfm?URI=oe-15-22-14870}

\bibitem{Grice2001Eliminating-frequency-and-space-time}
W.~P. Grice, A.~B. U'Ren, I.~A. Walmsley, Eliminating frequency and space-time
  correlations in multiphoton states, Physical Review A 64~(6) (2001) 063815.
\newblock \href {http://dx.doi.org/10.1103/PhysRevA.64.063815}
  {\path{doi:10.1103/PhysRevA.64.063815}}.

\bibitem{Mosley2008Conditional-preparation-of-single}
P.~J. Mosley, J.~S. Lundeen, B.~J. Smith, I.~A. Walmsley,
  \href{http://stacks.iop.org/1367-2630/10/093011}{Conditional preparation of
  single photons using parametric downconversion: a recipe for purity}, New
  Journal of Physics 10~(9) (2008) 093011 (29pp).
\newline\urlprefix\url{http://stacks.iop.org/1367-2630/10/093011}

\bibitem{Cohen2009Tailored-Photon-Pair-Generation}
O.~Cohen, J.~S. Lundeen, B.~J. Smith, G.~Puentes, P.~J. Mosley, I.~A. Walmsley,
  \href{http://link.aps.org/abstract/PRL/v102/e123603}{Tailored photon-pair
  generation in optical fibers}, Physical Review Letters 102~(12) (2009)
  123603.
\newblock \href {http://dx.doi.org/10.1103/PhysRevLett.102.123603}
  {\path{doi:10.1103/PhysRevLett.102.123603}}.
\newline\urlprefix\url{http://link.aps.org/abstract/PRL/v102/e123603}

\bibitem{Soller2011High-performance-single-photon-generation}
C.~S\"oller, O.~Cohen, B.~J. Smith, I.~A. Walmsley, C.~Silberhorn,
  High-performance single-photon generation with commercial-grade optical
  fiber, Phys. Rev. A 83~(3) (2011) 031806.
\newblock \href {http://dx.doi.org/10.1103/PhysRevA.83.031806}
  {\path{doi:10.1103/PhysRevA.83.031806}}.

\bibitem{Cui2012Minimizing-the-frequency-correlation}
L.~Cui, X.~Li, N.~Zhao,
  \href{http://stacks.iop.org/1367-2630/14/i=12/a=123001}{Minimizing the
  frequency correlation of photon pairs in photonic crystal fibers}, New
  Journal of Physics 14~(12) (2012) 123001.
\newline\urlprefix\url{http://stacks.iop.org/1367-2630/14/i=12/a=123001}

\bibitem{Yabuno2012Four-photon-quantum-interferometry}
M.~Yabuno, R.~Shimizu, Y.~Mitsumori, H.~Kosaka, K.~Edamatsu,
  \href{http://link.aps.org/doi/10.1103/PhysRevA.86.010302}{Four-photon quantum
  interferometry at a telecom wavelength}, Phys. Rev. A 86 (2012) 010302.
\newblock \href {http://dx.doi.org/10.1103/PhysRevA.86.010302}
  {\path{doi:10.1103/PhysRevA.86.010302}}.
\newline\urlprefix\url{http://link.aps.org/doi/10.1103/PhysRevA.86.010302}

\end{thebibliography}

\end{document}